\documentclass{aa}  

\usepackage{orcidlink}
\usepackage{graphicx}
\usepackage{hyperref}
\usepackage[dvipsnames]{xcolor}
\usepackage{txfonts}
\begin{document}

   \title{An oxygen-rich supernova remnant in NGC~253 with narrow [N\,II] emission lines}

   
   \author{T.~Kravtsov\inst{\ref{UTU},\ref{FINCA}}\orcidlink{0000-0003-0955-9102} \and
          E.~Congiu\inst{\ref{ESO}}\orcidlink{0000-0002-8549-4083} \and
          J.~P.~Anderson\inst{\ref{ESO}}\orcidlink{0000-0003-0227-3451} \and
          H.~Kuncarayakti\inst{\ref{UTU},\ref{FINCA}}\orcidlink{0000-0002-1132-1366} \and
          S.~Mattila\inst{\ref{UTU},\ref{Cyprus}}\orcidlink{0000-0001-7497-2994} \and
          A.~Amiri\inst{\ref{IPM},\ref{UA}}\orcidlink{0000-0002-8553-1964} \and
          A.~T.~Barnes\inst{\ref{ESOde}}\orcidlink{0000-0003-0410-4504}
          D.~A.~Dale\inst{\ref{UW}}\orcidlink{0000-0002-5782-9093} \and
          S.~Das\inst{\ref{SU},\ref{NHFP}}\orcidlink{0000-0002-9069-7061} \and 
          O.~V.~Egorov\inst{\ref{ARI}}\orcidlink{0000-0002-4755-118X}\and
          K.~Grasha\inst{\ref{ANU}}\orcidlink{0000-0002-3247-5321} \and
          E.~Habjan\inst{\ref{NU}}\orcidlink{0009-0003-9547-0952} \and
          S.~Mathur\inst{\ref{OSU}}\orcidlink{0000-0002-4822-3559} \and
          H.~Pan\inst{\ref{TU}}\orcidlink{0000-0002-1370-6964} \and
          T.~G.~Williams\inst{\ref{JBCA}}\orcidlink{0000-0002-0012-2142}
          }

   \institute{Tuorla Observatory, Department of Physics and Astronomy, University of Turku, 20014 Turku, Finland\label{UTU}\\ \email{thtkra@utu.fi}
   \and Finnish Centre for Astronomy with ESO (FINCA), University of Turku, 20014 Turku, Finland\label{FINCA}
   \and European Southern Observatory (ESO), Alonso de Córdova 3107, Casilla 19, Santiago 19001, Chile\label{ESO}
   \and School of Sciences, European University Cyprus, Diogenes Street, Engomi, 1516, Nicosia, Cyprus\label{Cyprus}
   \and School of Astronomy, Institute for Research in Fundamental Sciences (IPM), Tehran, P.O. Box 19395-5531, Iran\label{IPM}
   \and Department of Physics, University of Arkansas, 226 Physics Building, 825 West Dickson Street, Fayetteville, AR 72701, USA\label{UA}
   \and European Southern Observatory, Karl-Schwarzschild Straße 2, D-85748 Garching bei München, German\label{ESOde}
   \and Department of Physics and Astronomy, University of Wyoming, Laramie, WY 82071, USA\label{UW}
   \and Kavli Institute for Particle Astrophysics and Cosmology, Stanford University, 452 Lomita Mall, Stanford, CA 94305, USA\label{SU}
   \and NASA Hubble Fellow \label{NHFP}
   \and Astronomisches Rechen-Institut, Zentrum f\"{u}r Astronomie der Universit\"{a}t Heidelberg, M\"{o}nchhofstra\ss e 12-14, D-69120 Heidelberg, Germany\label{ARI}
   \and Research School of Astronomy and Astrophysics, Australian National University, Canberra, ACT 2611, Australia\label{ANU}
   \and Department of Physics, Northeastern University, 110 Forsyth St, Boston, MA 02115\label{NU}
   \and Astronomy Department, The Ohio State University, 140 West 18th Ave, Columbus, OH 43210, USA\label{OSU}
   \and Department of Physics, Tamkang University, No.151, Yingzhuan Road, Tamsui District, New Taipei City 251301, Taiwan\label{TU}
   \and UK ALMA Regional Centre Node, Jodrell Bank Centre for Astrophysics, Department of Physics and Astronomy, The University of Manchester, Oxford Road, Manchester M13 9PL, UK\label{JBCA}
             }

   \date{-}

   \titlerunning{O-rich SNR in NGC~253}
   \authorrunning{T.~Kravtsov, et al.}

 
  \abstract
   {Supernova remnants (SNRs) are the final evolutionary stage of supernovae. In rare cases, the remnant persists as an optically luminous remnant decades after the explosion. We report the discovery of an unusual oxygen-rich (O-rich) SNR that was serendipitously discovered in the MUSE mosaic of the nearby ($D \sim 3.5\,\mathrm{Mpc}$) galaxy NGC~253 released by \cite{Congiu2025}.}
   {We  search for SNRs in MUSE integral field spectroscopy data of NGC~253.}
   {We search for unresolved point sources with broad and strong [O\,III]$\lambda \lambda$4959,5007, [O\,I]$\lambda \lambda$6300,6364 or [O\,II]$\lambda \lambda$7320,7331 emission lines. We also search for possible counterparts at other wavelengths in archival data.}
   {We find a single O-rich SNR in NGC~253 with [O\,III]$\lambda \lambda$4959,5007. For this object, we report a coincident X-ray source detected by Chandra and a non-detection at mid-infrared wavelengths by JWST. Additionally, we identify [N\,II]$\lambda \lambda$6548,6583 emission lines and no detectable hydrogen or [SII]$\lambda \lambda$6716,6731 emission. The [N\,II] emission lines are broader (230 km/s) than the surrounding interstellar medium lines but still narrow in comparison to the [O\,III] emission (>1500 km/s) from the SNR. No detections of [O\,I]$\lambda \lambda$6300,6364 and [O\,II]$\lambda \lambda$7320,7331 were made.}
   {The discovery of a single O-rich SNR after a systematic search of NGC~253 highlights their scarcity in nearby galaxies. The structure of [O\,III] emission indicates complex structure in the O-rich ejecta. The relative narrowness of [N\,II] emission lines compared to the [O\,III] lines and the non-detection of hydrogen show that the remnant is surrounded by an N-rich circumstellar medium. The ejecta is currently interacting with this material of the preceding supernova explosion.}


   \maketitle
%

\section{Introduction}

A supernova remnant (SNR) is the final evolutionary phase of a supernova (SN) before it disperses into the interstellar medium. In the Milky Way, they are observed at a full complement of wavelengths from radio to gamma rays (e.g. \citealt{Ranasinghe2023, Vink2021}). In external galaxies, the radio emission often becomes too faint, and historically extragalactic SNRs have been identified via the use of optical emission lines, namely the ratio of [SII]$\lambda \lambda$6716,6731/H$\alpha$, which is enhanced in the shocked gas found in SNRs with respect to other nebulae such as H\,II regions \citep{Mathewson1972, Blair2004}. Recently, new techniques have been implemented for a more complete detection of SNRs. These include [OI]/H$\alpha$ line ratio mapping \citep{Kopsacheili2020}, which has been successfully implemented to detect hundreds of extragalactic SNRs by \cite{Li2024}. With the advent of high spatial resolution integral-field spectroscopy, shocked regions in galaxies, such as SNRs, may also be identified from their kinematics together with their spectral properties \citep{Congiu2023}. However, SNRs detected using these line-ratio methods are typically old, from at least thousands of years up to a million years \citep{Micelotta2018}.

Detected emission from SNRs is typically caused by the radiative cooling during a SNR's ``snowplough'' phase, where the ejecta and the ambient medium are piled into a dense cooling shell \citep{Micelotta2018}. There is usually a ``dark age'' period between the SN phase and its proceeding radiative SNR phase \citep{Yasuda2021}. For example, the initial phase of an H-rich Type~II SN is powered by recombination of hydrogen with possible additional energy coming from the ejecta interacting with the circumstellar medium (CSM). After recombination, the SN enters its nebular phase powered by the radioactive material produced by the explosion. As the radioactive material decays, the nebular emission weakens; the SN enters a freely expanding non-radiative phase for thousands of years, and observing it becomes a challenge. Finding young SNRs that are interacting and emitting light through other means is key to understanding their evolution during this ``dark age''.

Oxygen-rich (O-rich) SNRs are a rare and young subtype of SNRs, of which only a handful of examples are known; these include Cassiopeia A (Cas A), Puppis A in the Milky Way, LMC N132D in the Large Magellanic Cloud, and the luminous remnant SNR 4449-1 in NGC 4449 \citep{Blair2000, Ghavamian2023}. Their visible light spectra are dominated by the emission lines of oxygen originating from the inner parts of the remnant. In the case of SNR 4449-1, the reverse shock is thought to be interacting with the oxygen-rich layer within the ejecta originating from the carbon-burning process. The reverse shock itself originates from nitrogen-enriched CSM \citep{Milisavljevic2008}. Additionally, the fast shocks also produce extremely strong X-ray emissions in the ejecta \citep{Patnaude2003}. Age estimations of O-rich SNRs range from decades to a few thousand years, situating them between the nebular phase of a SN and the beginning of the snowplough phase of a SNR\citep{Mezcua2013, Bietenholz2010, Abdalla2021}.

Another unique SNR was recently discovered in M31 by \cite{Caldwell2023}, labeled WB92-26. The SNR exhibits strong oxygen emission lines in addition to [NII]$\lambda \lambda$6548,6583, [SII]$\lambda \lambda$6716,6731, and weak H$\alpha$ emission. The authors conclude that the strong nitrogen emission is caused by the CNO cycle enriching the outer layers of the progenitor star with nitrogen. However, it is still unclear if the emission originates from N-rich CSM or ejecta. 

Recently, several new O-rich SNRs were discovered in nearby galaxies using integral-field spectroscopy \citep{Kravtsov2025}. The method used in the discovery was based on the isolation of the wings of broad emission lines from the continuum and narrow emission lines. This allowed for the effective detection of SNRs, which tend to have broadened shock-excited emission lines compared to the narrow photoionised emission lines of H\,II regions. Motivated by these discoveries, in the current work we carried out a similar search in NGC~253 (the Sculptor Galaxy), which is a nearby galaxy with an accurately defined distance of 3.5 Mpc based on the tip of the red giant branch method (e.g.\ \citealt{Newman24, Radburn-Smith2011}). NGC~253 is a well-known starburst galaxy with numerous radio SNRs detected in the central 300 pc region \citep{Rampadarath2014}. These findings indicate a SN rate of up to 0.2 yr$^{-1}$, which is high compared to normal star-forming spiral galaxies.

In this article we report the discovery of a new oxygen-rich SNR in NGC~253. We describe the data and methods used in Section 2, present the results in Section 3, discuss the results in Section 4, and summarise the work in Section 5.

\section{Data and methods}

NGC\,253 was observed with the Multi Unit Spectroscopic Explorer (MUSE) on the Very Large Telescope (VLT) for over 50 hours (4$\times$211 s per pointing) to produce a $20^\prime \times 5^\prime$ mosaic of the whole galaxy. MUSE is a medium-resolution integral-field spectrograph (IFS) with a wavelength range of 4650-9350~\AA\,\citep{Bacon2010}. A full description of the data and reduction methods can be found in \cite{Congiu2025}.

The galaxy has also been observed with the James Webb Space Telescope (JWST) Mid-Infrared Instrument (MIRI) on November 4th 2023 with filters F770W and F2100W, and with the Hubble Space Telescope (HST) Advanced Camera for Surveys (ACS) on September 9 2006 with filters F475W, F606W and F814W. Data were downloaded from the Mikulski Archive for Space Telescopes\footnote{\hyperlink{https://mast.stsci.edu}{https://mast.stsci.edu}} as a Level 3 data product. The galaxy has additionally been observed with the Chandra X-ray Observatory and XMM-Newton at several epochs\citep{Lopez2023, Barnard2008}.

\begin{figure*}
   \centering
   \includegraphics[width=19cm]{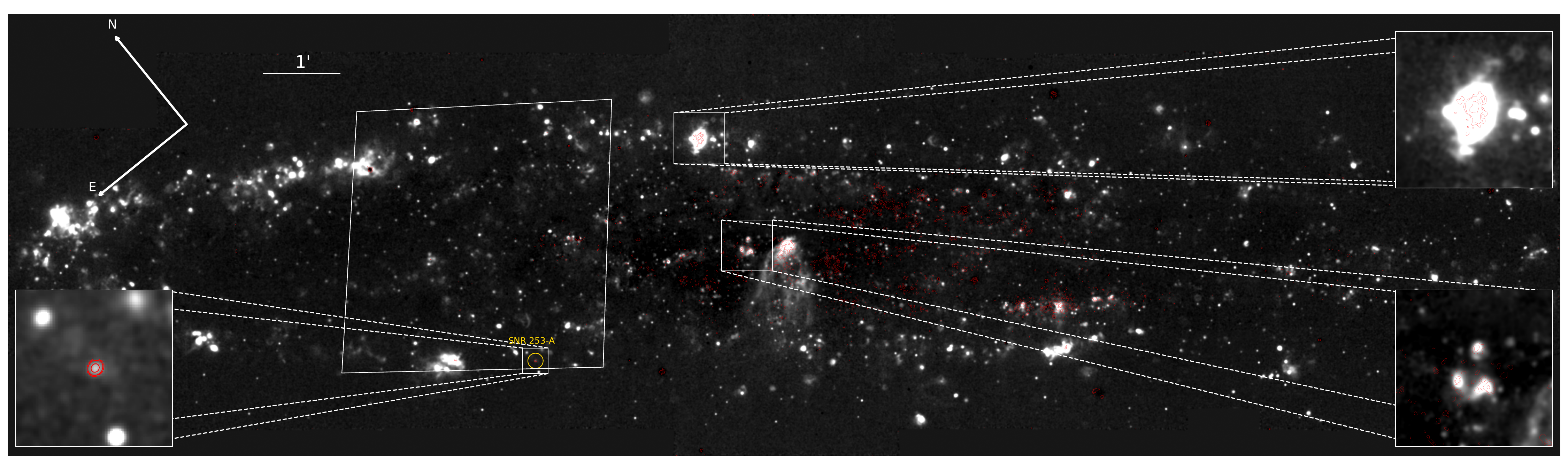}
      \caption{The grayscale image shows continuum-subtracted [O\,III]$\lambda$5007 linemap of NGC253. Bandpass of the linemap is 20 \AA\ with continuum samples taken from both sides from the emission line. Overlaid in red contours is the continuum-subtracted emission from between the [O\,III] doublet, 4965-5000 Å. The diffuse regions in red are mainly contamination from late-type stars in the galaxy, where absorption lines in the stellar continua sampled cause undersubtraction. Similarly, foreground stars with strong absorption features show up as bright unresolved sources. Some bright H\,II regions in NGC~253 show up as resolved regions in the contour map. The only O-rich SNR discovered in this paper is circled in yellow, and a zoomed-in region is shown in the inset (left). The unresolved nature of the source is evident when compared to other H\,II regions in the galaxy (insets on right). The large white parallelogram indicates the footprint of the broad-band HST/ACS  image shown in Figure \ref{hst_fov}. 
      }
         \label{ngc253_map}
   \end{figure*}

We searched for point sources with broadened emission lines of [O\,III]$\lambda \lambda$4959,5007, [O\,I]$\lambda \lambda$6300,6364 or [O\,II]$\lambda \lambda$7320,7331 employing the method outlined in \cite{Kravtsov2025}. To summarise, we generate continuum-subtracted linemaps from the MUSE observations that are offset by several angstroms from the observed wavelength of a given emission line. This way, narrowline emission sources are excluded from each linemap, which includes only broadline emission sources and contaminants. Contaminants include, e.g., foreground stars and bright H\,II regions, whose narrow but high S/N emission lines extend into the linemap region. A continuum-subtracted [O\,III]$\lambda$5007 linemap image along with a narrowband contour map of the continuum-subtracted emission between 4965 \AA \, and 5000 \AA \, is presented in Figure \ref{ngc253_map}. From these maps, we focused our search on emission sources that are unresolved and detectable in both the linemap and the contour map.

Oxygen lines are relatively weak in older SNRs dominated by interaction with the interstellar medium (ISM), which are typically brighter in [N\,II]$\lambda \lambda$6548,6583 and [S\,II]$\lambda \lambda$6716,6731. For possible counterparts of our optically uncovered O-rich SNRs, we investigated the emission locations using archival space telescope data, as these objects are known to have multiwavelength counterparts.

\section{Results}
\begin{figure}
   \centering
   \includegraphics[width=\hsize]{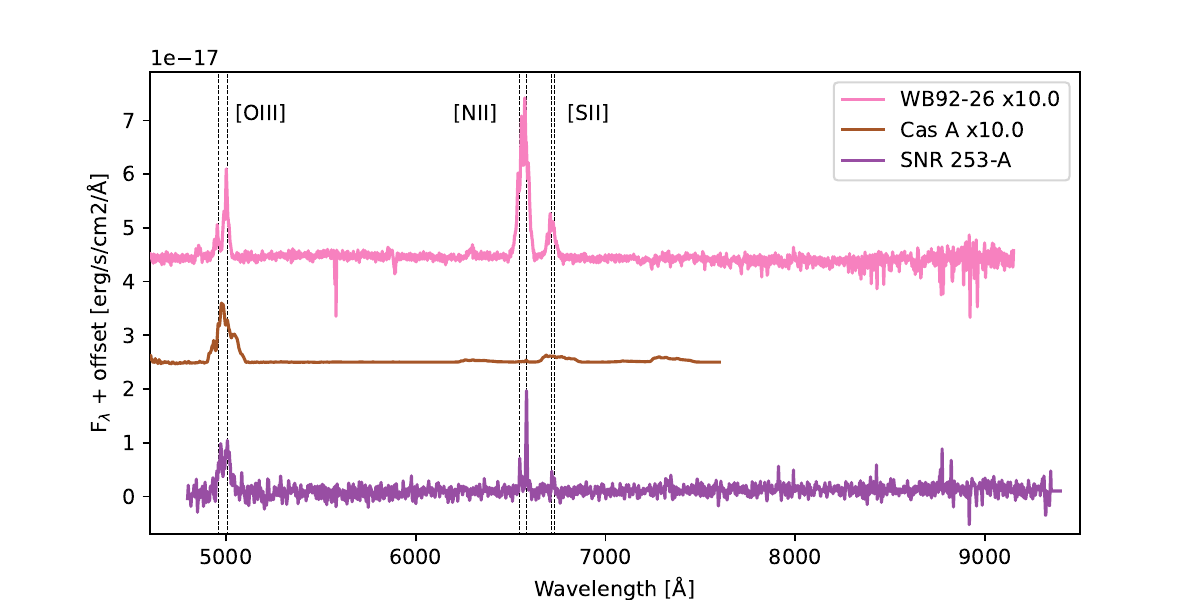}
      \caption{Extracted spectrum of the SNR discovered in NGC~253. The spectrum is compared to other known peculiar SNRs in the literature. The line fluxes have first been scaled to represent an observation at the same distance as NGC~253. The literature SNRs are fainter than the new SNR and have been scaled up by a factor of 10. Spectra are shifted vertically for clarity.
      }
         \label{spectral_comparison}
   \end{figure}

\begin{figure}
   \centering
   \includegraphics[width=\hsize]{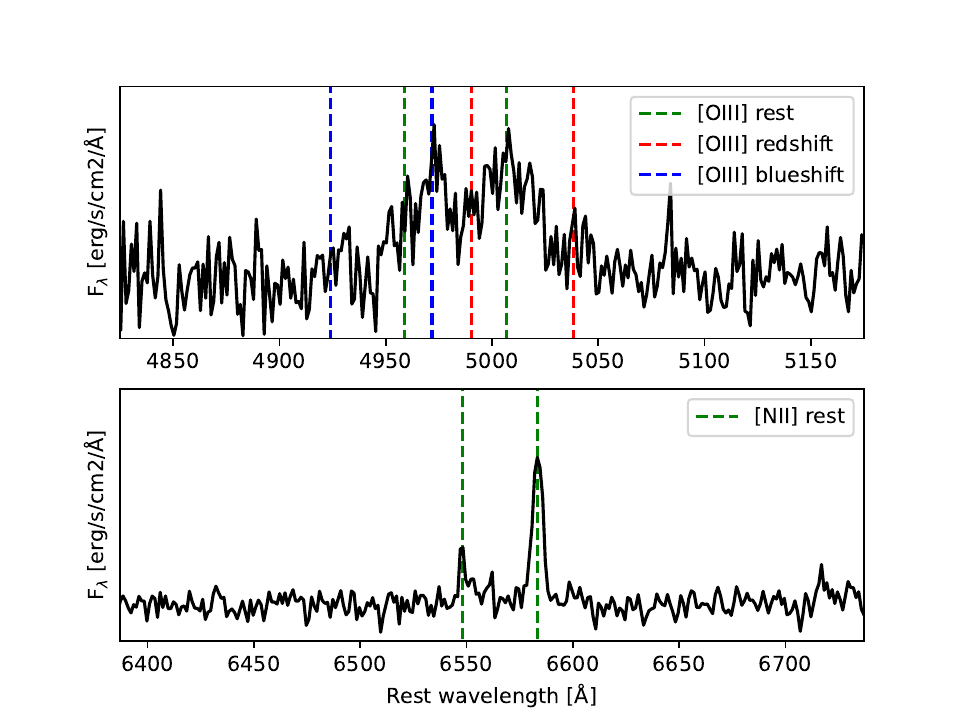}
      \caption{\textit{(Top)}: The spectra of [O\,III]$\lambda \lambda$4959,5007. The green dashed vertical line represents the oxygen doublet at rest. Redshift for the SNR was determined from a nearby H\,II region. If we assume that the highest peak is from [O\,III]$\lambda$5007, the bluer peak does not align with [O\,III]$\lambda$4959 at rest. Instead, it is more likely coming from blueshifted [O\,III]$\lambda$5007 (blue dashed line). There is an additional redshifted emission feature in the data, but it is unclear whether it is real or noise. \textit{(Bottom)}: The [N\,II]$\lambda \lambda$6548,6583 doublet of the SNR. There is no detectable H$\alpha$ emission from the source.
              }
         \label{spectral_zoom_box}
   \end{figure}

We find one SNR with clearly detectable and broad oxygen emission lines within the [O\,III]$\lambda \lambda$4959,5007 linemap. However, we do not find any broadline SNRs within the [O\,I]$\lambda \lambda$6300,6364 and [O\,II]$\lambda \lambda$7320,7331 linemaps. We adopt coordinates RA=00:47:49.05 Dec=-25:16:28.7 (J2000.0) for the location of the remnant based on its [O\,III]$\lambda$5007 centroid and label it as SNR 253-A. We extracted the full spectrum of the source with a circular aperture (radius of 1") from the MUSE datacube and subtracted the background emission with an 0.2" annular ring around the aperture (Fig \ref{spectral_comparison}). The extracted spectrum shows only emission of [O\,III]$\lambda$5007 and [N\,II]$\lambda \lambda$6548,6583 (Fig. \ref{spectral_zoom_box}). The oxygen emission line itself shows three separate components, which we interpret as the $\lambda$5007 line at three different velocities. We measure the two stronger emission features to have full-width at half maximum of 1570 km/s and 1760 km/s by fitting two Gaussian profiles to them. The doublet counterpart, $\lambda$4959, is not detected. The doublet has an intrinsic line ratio of 3:1 and this is therefore explained by noise. The [N\,II]$\lambda \lambda$6548,6583 doublet is broadened to a velocity of 229 km/s after removing the instrumental broadening.

\begin{figure*}
        \centering
        \includegraphics[width=\textwidth]{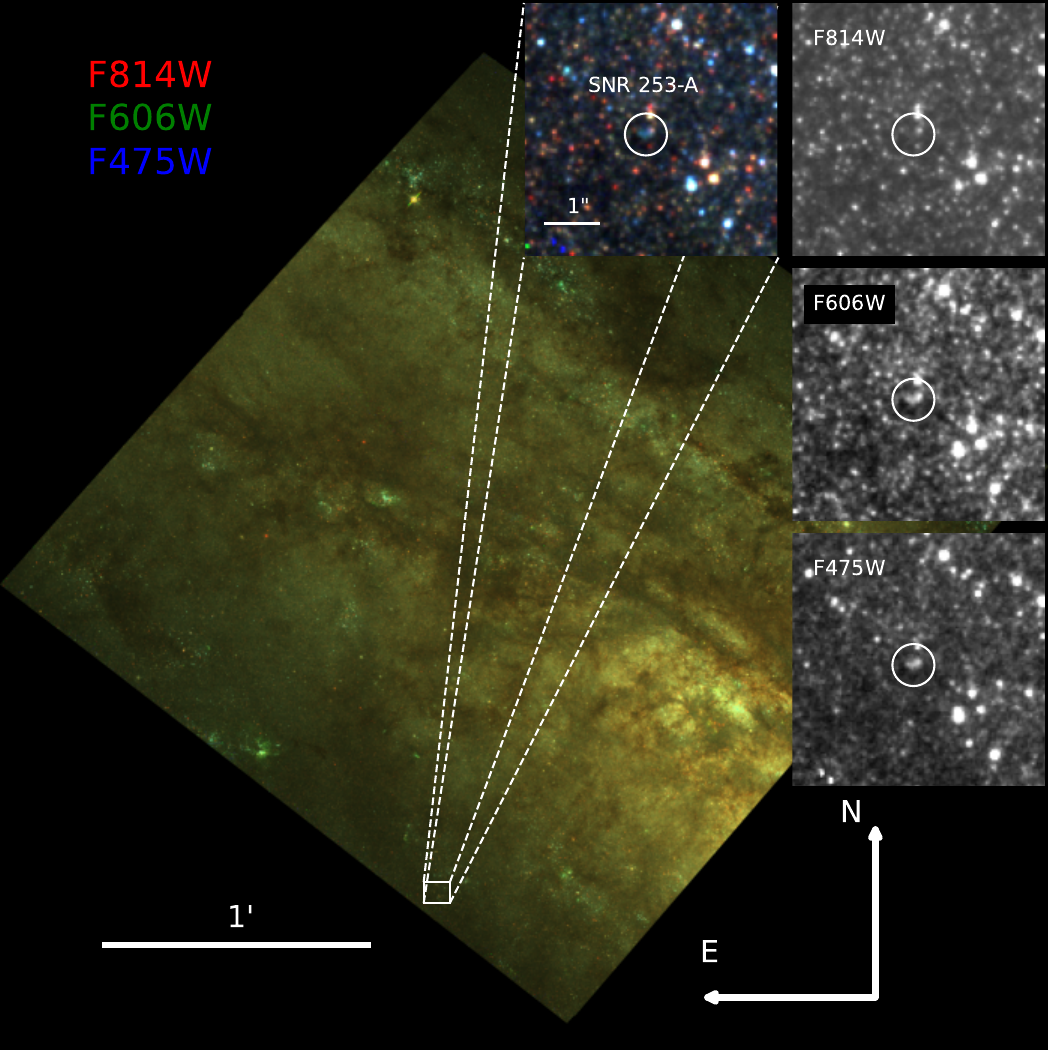}
        \caption{Archival HST/ACS images of a sub-region of NGC~253 in filters F475W, F606W and F814. The zoomed-in insets of the O-rich SNR region (white square) are presented on the right. There is a blue-green peanut-shaped source (white circle) near the center of the location, and both MUSE and Chandra detections are within 1" of this source. Black and white images show the non-detection in the F814W band. The F475W and F606W bandwidths overlap with [OIII]$\lambda \lambda$4959,5007 and [NII]$\lambda \lambda$6548,6583, respectively, indicating that this is likely the source of the spectrum.
          }
        \label{hst_fov}
   \end{figure*}

Searching for the source in space telescope images, we find a resolved lobe-like structure visible in two HST/ACS imagess, F475W and F606W (Figure \ref{hst_fov}). Considering that these filters coincide with the [O\,III]$\lambda\lambda$5007,4959 and [N\,II]$\lambda \lambda$6548,6583 emission, and that the structure is not detected in the third filter, F814W, we identify this structure as the source of the emission. However, in all images there is an unresolved reddish feature overlapping with the structure. The field is dense with old red supergiants, and we identify this feature to be a nearby star unrelated to the SNR. Based on the two HST images, the size of the source is 0.40" $\times$ 0.34" (cf. typical full width at half maximum of 0.08"), corresponding to a physical size of 6.8 pc $\times$ 5.8 pc \cite[cf. Cassiopeia A's diameter of 5 pc from e.g.][]{DeLooze2017}. Assuming an expansion velocity of 6000 km/s, similar to the highest velocities in Cas A's emission shell \citep{Fesen2025}, this size would indicate an age of $\sim$500 years. 

From the Chandra Source Catalog\footnote{\hyperlink{https://cxc.cfa.harvard.edu/csc/}{https://cxc.cfa.harvard.edu/csc/}}, we find a source 2CXO J004749.0-251628 with sub-arcsecond separation from the HST detection. The source has a measured flux of 2.8$_{-0.9}^{+0.8}$ $\times$ 10$^{-15}$ erg s$^{-1}$ cm$^{-2}$, corresponding to an X-ray luminosity of 4.1$_{-1.3}^{+1.2}$ $\times$ 10$^{36}$ erg s$^{-1}$.  We also note a non-detection in the JWST images with filters F770W and F2100W and in the XMM-Newton observations.

\section{Discussion}
We have employed a recently presented method to find extragalactic SNRs in the nearby starburst galaxy NGC~253. The method exploits the field of view and spatial resolution of the MUSE IFU. The observations cover most of the galactic disk of NGC~253 with over a hundred pointings covering over 100 arcmin$^2$ on the sky.

We detected only one O-rich SNR in NGC~253 with the MUSE datasets. With the aid of archival HST and Chandra observations, we found a shell-like structure and an X-ray point source coinciding with the MUSE detection. In addition, we found no signal in JWST images with filters F770W and F2100W that would originate from a compact source. This is in contrast to other recently discovered O-rich SNRs, most of which exhibit moderately strong mid-IR emission.

\subsection{Counterparts from other telescopes}
We detected a small resolved counterpart in two of the three filters available in the HST archival data. Based on the filter bandpasses, we identify these to be [O\,III]$ \lambda \lambda$4959,5007 and [N\,II]$\lambda \lambda$6548,6583 in F475W and F606W, respectively. The two emitting regions overlap each other considerably, and we detect no significant difference between the sizes of the regions. In the Chandra archival data, we found a point source that coincides with the HST counterpart within Chandra's pointing accuracy. The measured X-ray luminosity of 4.1 $\times$ 10$^{36}$ erg s$^{-1}$ is of the same order as that for SNRs found in the LMC \citep{Ou2018}. This is roughly one third of the X-ray luminosity of Cas A, 1.35 $\times$ 10$^{37}$ erg s$^{-1}$ \citep{Arias2018}. Conversely, the [O\,III]$ \lambda \lambda$4959,5007 luminosity of SNR 253-A is a factor of five larger than that of Cas A (Fig. \ref{spectral_comparison}). It is unclear what is causing the differences between these SNRs, but it may be intrinsic to each SNR and their environments. For example, the CSM structure may be different, which leads to different shock interaction and thermal X-ray emissions. The oxygen yields from the SNe may also have been different which may lead to different spectral properties within the inner ejecta. Age may also play a role as the derived of Cas A is approximately 350 years, which is about 150 years younger than the age of 500 years estimated for SNR 253-A \citep{Fesen2025}.

The non-detection from JWST puts an upper limit on the dust present at the SNR site. We measure upper limits of 7.3~$\mu$Jy and 49.2~$\mu$Jy in the two filters F770W and F2100W, respectively. For simplicity, we assume a Milky Way dust composition and adopt a cool dust temperature of 120 K, based on the dust models previously used for SNe (e.g., SN\,1993J by \citealt{Szalai2025}). With these assumptions and while only two filters were used, the detection limits place the dust mass at $\lesssim$10$^{-4}$ M$_\odot$. This makes the SNR extremely dust-poor in comparison to other SNRs of similar ages, where the dust masses have been estimated to be between 0.1-1 M$_\odot$\citep{DeLooze2017, Niculescu-Duvaz2022}. Other O-rich SNRs have been shown to exhibit mid-IR emission, so it is unclear what is causing the lack of dust emission \citep{Milisavljevic2024, Kravtsov2025}. The non-detection may be an indication that the amount of dust created after the SN is very low, or that any dust present has actively been destroyed by interaction between SN ejecta and the surrounding material. Alternatively, the dust may be extremely cool and not detectable in these mid-IR observations. It may be that some SNe intrinsically produce more dust or that they are more efficient dust destroyers. The small sample of O-rich SNRs happens to include mostly dust rich objects with large mid-IR excesses. While in the case of this SNR we cannot distinguish between these scenarios with the current dataset, future deeper IR observations may reveal the presence of dust or constrain the dust mass more robustly.

\subsection{Observed emission lines}
The discovered source only shows emission lines of [O\,III]$ \lambda$5007 and [N\,II]$\lambda \lambda$6548,6583 in the MUSE spectrum. While the [O\,III]$ \lambda \lambda$4959,5007 region shows complex structure with several peaks, we identified them all to arise from [O\,III]$ \lambda$5007 with different velocities for two reasons. First, the separation between the two main peaks is measurably shorter in wavelength than the intrinsic separation between the doublet components. Second, the intrinsic flux ratio of the double is a constant $\sim$3:1, while the observed ratio is closer to 1:1. We therefore conclude that most of the detected oxygen spectrum is likely from [O\,III]$ \lambda$5007, while [O\,III]$ \lambda$4959 is buried in the noise.

If we assume that the elements in a SNR are thoroughly mixed and are shocked by the same process, we would expect to see similar spectral line profiles emerge from the shocked gas. Therefore, the observed discrepancy in linewidths between the oxygen and nitrogen emission, $\sim$1650 km/s and $\sim$230 km/s, respectively, suggests that they originate from different locations in the remnant. While the oxygen emission is complicated and likely filamented, the nitrogen emission lines are symmetrical and only mildly broadened. The nitrogen is therefore likely related to the ejecta-CSM interaction, where the slow-moving CSM is enriched with nitrogen from the CNO processes of the progenitor. The X-ray detection may then be explained by the hot gases produced in the interaction shocks. This is akin to more typical SNRs found in nearby galaxies with X-ray observatories. The differences in linewidths between the two elements indicate that the emitting region is highly complex, with several different regions moving at different peculiar velocities. This may be an indication that the oxygen emission is related to filamented inner parts of the fast-moving ejecta being shocked. Strongly filamented structure within the ejecta would then explain the observed structure of the emission line, and similar multicomponent features are observed in other extragalactic O-rich SNRs \citep{Milisavljevic2008}.

We also note the absence of H$\alpha$ and [S\,II]$\lambda \lambda$6716,6731 emission lines, which are typically detected in SNRs. The missing sulfur emission may be explained by low signal-to-noise or densities higher than the critical density of these forbidden lines. This still leaves the question of missing hydrogen, which would be expected to be ionised in the shocked region with H$\alpha$ intensity of similar levels as nitrogen. It may therefore be possible that the observations are caused by an overabundance of nitrogen in the interacting material, greatly enhancing the [N\,II]$\lambda \lambda$6548,6583 emission intensity compared to H$\alpha$. $\eta$ Carinae is known to have nitrogen-rich ejecta surrounding it from previous eruptions, and this SNR may therefore be embedded inside a similar structure \citep{Mehner2016}. Similarly, a post-merger binary star system HD 148937 in the Milky Way has a CNO-rich nebula surrounding it \citep{Frost2024, Mahy2017}. Here the star system is surroundend by a nebula, NGC 6164/65, which is overabundant in nitrogen. In any case, deeper observations will be needed in order to determine if this O-rich SNR is an older remnant starting to exhibit narrow typical SNR features or if the apparent nitrogen enrichment is real.

\section{Summary}
This work aimed to search NGC~253 for O-rich SNRs. We utilised the extensive MUSE observations of the whole galaxy, comprising of over 100 individual pointsings totaling over 50 hours of total exposure time, searching for strong and broad oxygen emission sources. Only one such source was discovered in the galaxy. In addition, we find a resolved shell-like SNR in archival HST images and an X-ray point source coinciding with the SNR location.

The complex structure of the oxygen emission feature originates from the filamentary inner ejecta, while the simple nitrogen emission is indicative of shocked interaction with the CSM. This, and the resolved structure of several parsecs, indicate that the SNR is hundreds of years old. If the oxygen emission is powered by the reverse shock of the explosion, the interaction should be weakening at this point, and the O-rich features should disappear in the coming years. This will give an excellent opportunity to study late-stage evolution of O-rich SNRs.

For future work, we will aim to investigate more galaxies in the Local Volume as they get observed with MUSE. It may also be possible to expand the search for O-rich SNRs to the northern sky with other telescopes and the careful selection of narrowband or tunable filters. Expanding the catalogue of O-rich SNRs is critical when investigating their prevalence, origins and powering mechanisms.

\begin{acknowledgements}
    We thank Dan Milisavljevic for the integrated spectrum of Cas A. T.K. thanks Peter Lundqvist for fruitful discussions on SNR evolution. Based on observations collected at the European Organisation for Astronomical Research in the Southern Hemisphere under ESO programmes 108.2289 (PI: Congiu) and 0102.B-0078(A) (PI: Zschaechner). H.K. was funded by the Research Council of Finland projects 358691 and 358692. S.M. acknowledges financial support from the Research Council of Finland project 350458. S.D. is grateful for support from NASA through the NASA Hubble Fellowship grant HST-HF2-51551.001-A, awarded by the Space Telescope Science Institute, operated by the Association of Universities for Research in Astronomy, Incorporated, under NASA contract NAS5-26555. OE acknowledges funding from the Deutsche Forschungsgemeinschaft (DFG, German Research Foundation) -- project-ID 541068876. HAP acknowledges support from the National Science and Technology Council of Taiwan under grant 110-2112-M-032-020-MY3 and 113-2112-M-032-014-MY3. TGW gratefully acknowledges support from the UK ALMA Regional Centre (ARC) Node, which is supported by the Science and Technology Facilities Council grant number ST/Y004108/1.
\end{acknowledgements}

\bibliographystyle{aa}
\bibliography{letter25}

\end{document}